\documentclass[fleqn,10pt,twocolumn]{wlscirep}

\usepackage{pgfplots}
\usepackage[export]{adjustbox}
\usepackage{caption}
\usepackage{subcaption}
\usepackage{lineno}
\usepackage{amsmath}
\usepackage{amssymb}
\usepackage{graphicx}
\usepackage{microtype}
\usepackage{times}

\usepackage[font=footnotesize,labelfont=bf,
   justification=justified,
   format=plain]{caption} 

\pgfplotsset{compat=1.12}

\title{Targeted Nuclear Spallation from Moderator Block Design for a Ground-Based Space Radiation Analog}

\author[1,*]{Jeffery C.~Chancellor}
\author[2]{Stephen B.~Guetersloh}
\author[3]{Rebecca S.~Blue}
\author[4]{Keith A.~Cengel}
\author[5]{John R.~Ford}
\author[1,6,7]{Helmut G.~Katzgraber}
\affil[1]{Department of Physics and Astronomy, Texas A\&M University, College Station, Texas 77843-4242, USA}
\affil[2]{Practical Physics Solutions, LLC, College Station, 77843-4242, USA}
\affil[3]{Aerospace Medicine and Vestibular Research Laboratory, The Mayo Clinic Arizona, Scottsdale, AZ 85054, USA}
\affil[4]{Perlman School of Medicine, University of Pennsylvania, Philidelphia, USA}
\affil[5]{Department of Nuclear Engineering, Texas A\&M University, College Station, 77843-4242, USA}
\affil[6] {Santa Fe Institute, 1399 Hyde Park Road, Santa Fe, New
Mexico 87501 USA}
\affil[7]{Microsoft Quantum, Microsoft, Redmond, Washington 98052, USA}

\affil[*]{jeff@chancellor.space}

\keywords{radiation, spallation, metamaterials}

\begin{abstract}
Current radiobiology studies on the effects of galactic cosmic ray radiation utilize mono-energetic beams, where the projected dose for an exploration mission is given using highly-acute exposures. This methodology does not replicate the multi-ion species and energies found in the space radiation environment, nor does it reflect the low dose-rate found in interplanetary space. In radiation biology studies as well as in the assessment of health risk to astronaut crews, the differences in the biological effectiveness of different ions is primarily attributed to differences in the linear energy transfer (LET) of the radiation spectrum. Here we show that the LET spectrum of the intravehicular environment of spaceflight vehicles can be simulated with a single particle, mono-energetic ion beam accelerated at target blocks constructed of one or more materials. The LET spectrum of the emerging field can then be moderated by the amount of mass or length of material the primary and secondary nuclei travels, thus preferentially producing specific nuclear spallation and fragmentation processes and allowing for a continuous generation of ionizing radiation that mimics the space radiation environment. This approach could allow more accurate simulation of not only intravehicular spaceflight conditions, but also could be used to simulate the external galactic cosmic ray field, planetary surface spectrum (e.g., Mars or Moon), and the local radiation environment of orbiting satellites, providing a much-needed ground-based space radiation analog for future experimentation.
\end{abstract}
\begin{document}

\flushbottom
\maketitle

\section*{Introduction} \label{sec:intro}

During spaceflight, astronauts are exposed to a variety of environmental stressors ranging from chemical and bacterial insults (resulting from the materials and occupants of the space vehicle) to microgravity and mixed fields of ionizing radiation. The space radiation environment is a complex combination of fast-moving ions derived from all atomic species found in the periodic table, with any meaningful abundance up to approximately nickel (atomic number $Z=28$). These ionized nuclei have sufficient energy to penetrate the spacecraft structure and cause deleterious biological damage to astronaut crews and other biological material, such as cell and tissue cultures \cite{Chancellor,Walker}. Recent studies have demonstrated that the biological response and disease pathogenesis to space radiation is unique to a nonhomogeneous, multi-energetic dose distribution similar to the interplanetary space environment \cite{Kennedy,RomeroWeaver2014}. Previous radiobiological models and experiments utilizing mono-energetic beams may not have fully characterized the biological responses or described the impact of space radiation on the health of vital tissues and organ systems \cite{Chancellor_2018}.

Currently, radiobiology studies on the effects of {\em galactic cosmic ray} (GCR) radiation utilize single ion, mono-energetic beams (e.g., Li, C, O, Si, Fe, etc.) at heavy-ion accelerators where the projected dose for an entire exploration-class mission is given to biological models using highly-acute, single ion exposures. Recently, a GCR simulator was developed that can provide three to five consecutive mono-energetic heavy ions for space radiation studies \cite{hidding2017laser,norbury2016galactic}. While an improvement upon previous capabilities, this approach only provides a few data points and lacks the generation of pions and neutrons that account for as much as $15$-$20$\% of a dose exposure \cite{Slaba2015}. Additionally, for radiobiology studies, questions still remain regarding what order the ion species should be given as this may affect experimental outcomes \cite{Elmore2011,Fry2002}.

Unfortunately, these approaches do not reflect the low dose-rate found in interplanetary space, nor do they accurately replicate the multi-ion species and energies found in the GCR radiation environment. It is believed that the complex GCR environment could cause multi-organ dose toxicity, inhibiting cell regrowth and tissue repair mechanisms. Thus, high-fidelity simulation is critical for the determination of accurate radiobiological experimental outcomes \cite{Chancellor_2018,wilson1995issues}. Furthermore, interaction with the spacecraft hull attenuates the energy of heavy charged particles and frequently causes their fragmentation into lighter, less energetic elements, changing the complexity and makeup of the {\em intravehicular} (IVA) radiation spectrum. Therefore, it is important for the fidelity of space radiation studies that the space radiation environment both outside and inside spacecrafts can be accurately simulated.

\begin{figure}[ht]
\centering
\includegraphics[scale=0.40]{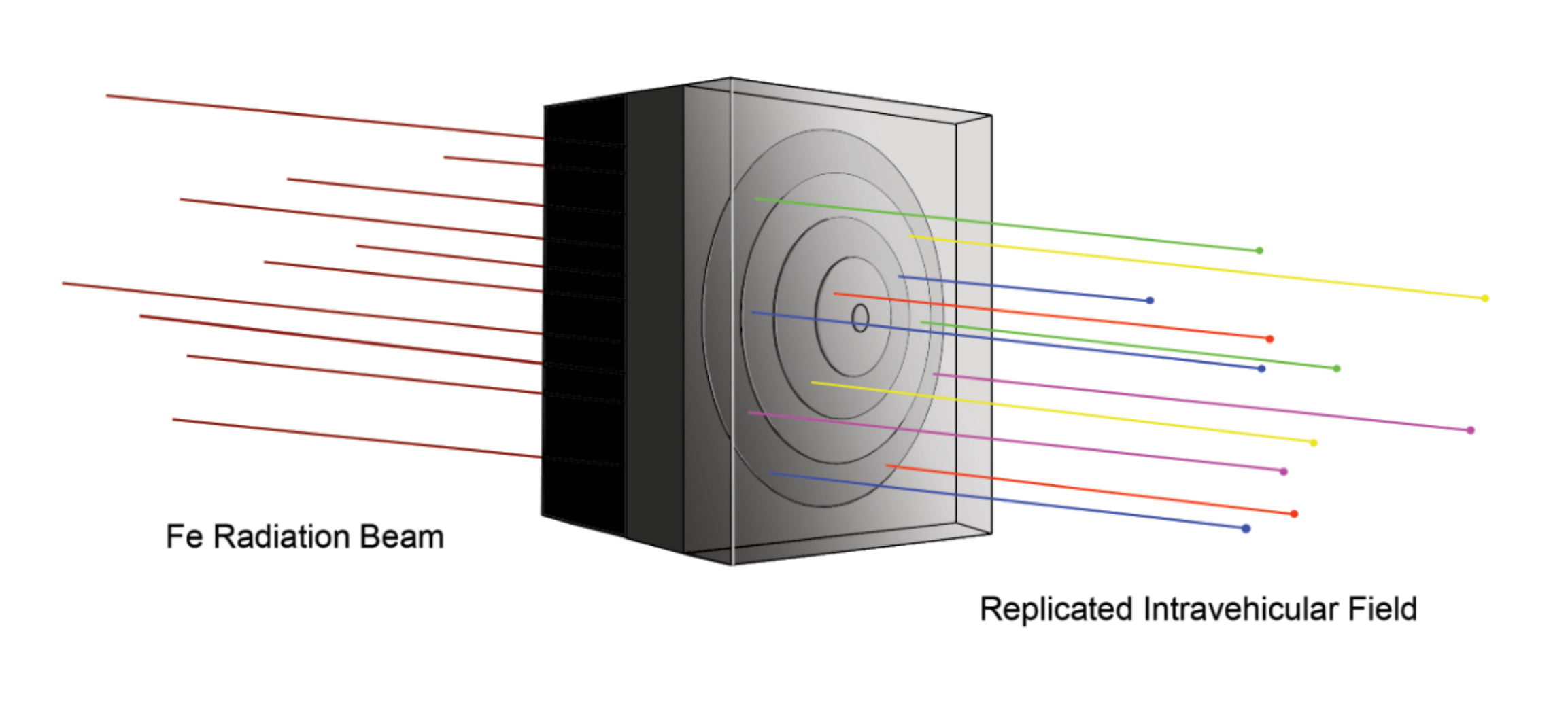}
\caption{Moderator block geometry concept for the emulation of space radiation spectra. A primary beam of $^{56}$Fe (iron, left) is selectively degraded with a carefully designed moderator block to produce a desired distribution of energies and ions (represented by the colorful lines on the right) simulating the intravehicular space radiation environment. Figure reprinted with permission from Chancellor et al.\cite{Chancellor_2018}, under the \href{https://creativecommons.org/licenses/by/4.0/}{Creative Commons license}.}
\label{fig:blockconcept}
\end{figure}

In this work we demonstrate an approach to simulate the space radiation environment in a laboratory setting. For simplicity, we focus on the IVA radiation spectrum measured on different spacecrafts. Our goal is to numerically develop a target moderator block that can be easily constructed from materials with multiple layers of varying geometry to generate specific nuclear reactions and spallation products. The moderator block is designed so that the final field closely simulates the IVA \emph{linear energy transfer} (LET) spectrum measured on previous spaceflights. The LET quantifies how much energy is lost in a material and is typically given in units of kilo electron volts per micron (keV/$\mu$m) for quantification of radiobiological damage. This proposed target moderator block can, for example, be placed in front of a $1$ giga electron volt per nucleon (GeV/n) iron ($^{56}$Fe) single-particle beam with no modifications to the beamline infrastructure. As the iron beam passes through the moderator block, nuclear spallation processes can create modest amounts of the desired fragments resulting in a complex mixed field of particle nuclei with different atomic numbers $Z$ in the range $0< Z \leq 26$ and LETs up to approximately 200 keV/$\mu$m. Modifications to the internal geometry and chemical composition of the materials in the target moderator block allow for a shaping of the simulated IVA LET to specific spectra. The concept is shown in Figure \ref{fig:blockconcept}. Our approach thus leverages available beamline technologies to provide an enhancement to current ground-based analogs of the space radiation environment by reproducing the measured IVA LET spectrum.  

To demonstrate applicability of this approach, results from our numerical models are compared below to real-world measurements of the IVA LET spectrum from the U.S. Space Shuttle, International Space Station (ISS), and  and NASA’s new Orion Multi-Purpose Crew Vehicle (MPCV). While these intravehicular environments were chosen for concept demonstration, we emphasize that this approach could be generalized to other radiation spectra and a wide range of environmental conditions for radiobiological studies as well as other applications such as the testing of shielding, electronics, and materials for a space environment, or for nuclear research facilities and laboratories.

\section*{Background} \label{sec:background}

Highly-charged heavy ions penetrate matter with an approximate straight path. The interaction of the ion with material's atomic structure results in one of two outcomes: transfer of energy from the primary ion into the medium (gradually dissipating the primary ion's energy) or the creation of progeny nuclei and spallation fragments \cite{rutherford2012scattering}.

The loss of an ion's energy can be accurately approximated using the stopping power equation, \cite{Fano1963}

\begin{eqnarray}\label{eq:BB} \frac{dE}{dx} = \frac{4\pi e^{4}Z_{1}^{2}Z_{2}}{m_{e}\beta^{2}} &&
\!\!\!\!\!\!\! \Big[ \ln\left(\frac{2m_{e}v^{2}}{I}\right) - \Big. \\ \nonumber && - \Big.
\ln(1-\beta^{2})-\beta^{2} - \frac{C}{Z_{2}} - \frac{\delta}{2} \Big]\, .
\end{eqnarray}

\noindent Here, $Z_{1}$ and $Z_{2}$ are the charges of the primary ion and the medium being traversed, respectively; $m_{e}$ is the electron mass density of the medium and $\beta = v/c$ with $v$ representing the velocity of the primary ion and $c$ representing the speed of light. Material-specific effects are described by the average ionizing potential of the medium $I$, the shell correction $C/Z_{2}$, and the density effect $\delta$. These relationships have been validated with both theoretical and experimental results.\cite{bischel1964,national1964studies,Sternheimer1982,Ziegler2010}

Analysis of Eq. \eqref{eq:BB} shows that a charged particle traversing a given material will lose kinetic energy at a rate inversely proportional to its speed, with a prompt loss of energy as it comes to rest. This sudden rise in energy loss is referred to as the \textit{Bragg peak}. In this context, the term "Bragg peak" differs from the definition used in materials and condensed matter studies. In radiation dosimetry, reference to the Bragg peak implies the point at which a charged particle promptly loses kinetic energy before coming to rest in a medium.

As mentioned above, there is a chance that the interaction between the primary and the nuclear structure results in the dislocation of nuclear matter from the primary ion, creating fragments of ion species each with charges equal to or less than the charge of the primary ion. Brandt and Peters demonstrated that the cross-section for a nuclear interaction to induce a charge-changing spallation can be determined with,\cite{bradt1950heavy,wilson1986}

\begin{equation}\label{eq:bradtandpeters}
\sigma = \pi r_{0}^{2} \Big[\sqrt[3]{A_{\rm P}} + \sqrt[3]{A_{\rm T}} - \delta (A_{\rm T},A_{\rm P},E)\Big]^{2},
\end{equation}

\noindent where $A_{\rm P}$ and $A_{\rm T}$  are the mass numbers of the primary ion and target medium, respectively, $\delta $ is a fitted parameter dependent on the energy of the primary ion, and $r_{0} = 1.26$~fm.

These phenomena described in Eqs. \ref{eq:BB} and \ref{eq:bradtandpeters} provide valuable information about the character and properties of materials, and provide a novel method of generating a mixed field of ions using accelerator technologies. Careful observation of Eq. \eqref{eq:BB} shows that a primary ion can penetrate a material of thickness, $\Delta x$, assuming a high enough incident energy, $E$. We can choose $\Delta x$ so that roughly half of the primary ions have a charge-changing reaction defined by Eq. \eqref{eq:bradtandpeters}. With these constraints the location of the first charge-changing interaction will occur, on average, at approximately the midpoint (e.g., $\Delta x /2$). Most nuclear reactions are peripheral and remove only a few nucleons from the incident primary ion. The resulting secondary particles will then travel a distance equal to approximately half of the material's thickness, $\Delta x$, and will lose energy at a rate described by Eq. \eqref{eq:BB}. On average, observing Eq. \eqref{eq:bradtandpeters}, the probability of the lighter fragments having tertiary interactions is proportional to their atomic mass, e.g. $~A^{2/3}$, or approximately 25\%. Thus, for a primary ion of constant energy $E$, atomic mass $A_{\rm P}$, and charge $Z$, incident on a target with atomic mass $A_{\rm T}$ and thickness $\Delta x$, the emerging field will consist of a mixed ion species with varied charges up to the primary ion's atomic mass, $A_{\rm P}$.

The stopping power described in Eq.~\ref{eq:BB} is equivalent to the energy loss per unit path length of the primary ion, or LET, thus ${\rm LET} = dE/dx$, and quantifies how much energy is lost in a material. Note, here we infer the \emph{unrestricted stopping power}, e.g. all the energy loss by the primary is into the medium. LET is typically given in units of mega electron volts per centimeter (MeV/cm) for materials studies; however, for radiobiological quantification where the outcome varies at distances of 10$^{-6}$~m, LET is given in keV/$\mu$m. Although not uniquely related to biological response, LET is an important metric that is utilized to determine radiation tissue damage where the differences in the {\em relative biological effectiveness} (RBE) of different ions are, in part, attributed to differences in the LET of the radiation \cite{icrp60}. The RBE of a particular radiation type is the numerical expression of the relative amount of damage that a fixed dose of that type of radiation will have on biological tissues. LET remains the focus of many biological investigations and serves as the basis of radiation protection and risk assessment \cite{wilson1995issues,icrp60}.

\begin{figure}[ht]
\centering
\includegraphics[width=0.95\columnwidth]{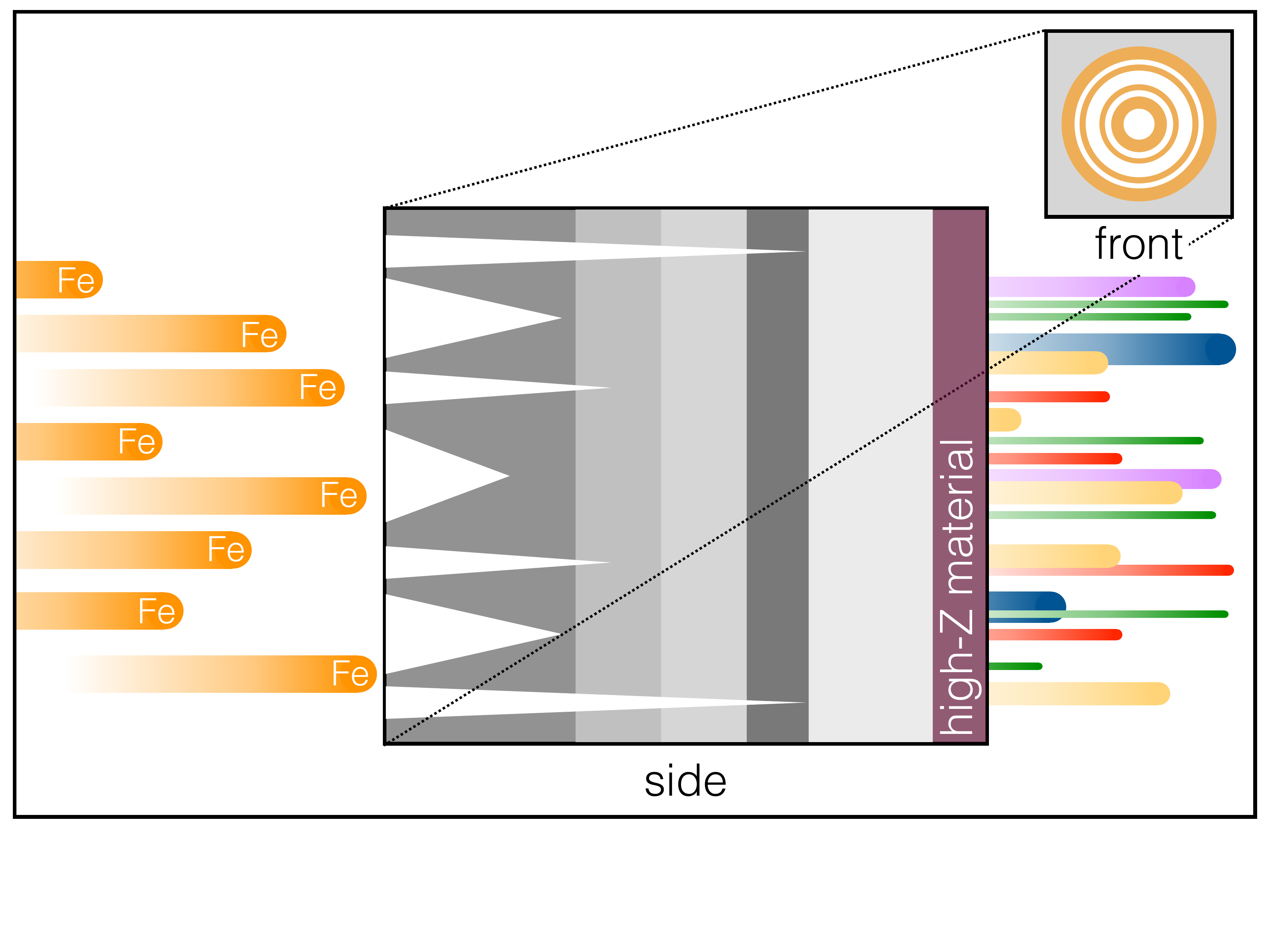} \vspace*{-2.5em}
\caption{ Schematic of the moderator block designed to simulate specific space radiation spectra. A primary beam of $^{56}$Fe (left) is selectively degraded with a carefully-designed moderator block to produce a desired distribution of energies and ions (represented by the colorful lines on the right-hand side). To preferentially enhance fragmentation and energy loss, cuts (white sections on the left-hand side) are performed in the moderator block made up of different materials (depicted by different shades of gray). Before the spallation products exit the moderator block, a high-$Z$ material layer is added for scattering. The inset shows the circular beam spot, as well as the symmetric cuts made into the moderator block.}
\label{fig:schematic}
\end{figure}

Conceptually, it is reasonable to predict that a single particle, mono-energetic ion beam can be accelerated at target blocks constructed of one or more materials. The spectrum of the emerging field can then be moderated by the amount of mass or length of material the primary and secondary nuclei travels. The robustness of the resulting field of mixed ions and energies would be dependent on the careful selection of target material(s) and the relative contribution of each layer to the desired spectrum. Figure \ref{fig:schematic} demonstrates this concept.

\section*{Methods}\label{sec:methods}
Highly-charged heavy ions penetrate a material with an approximate straight path and gradually dissipate energy through multiple collisions with the material's electronic structure. Eq.~\ref{eq:bradtandpeters} shows that the effectiveness of a material to instigate energy loss attenuation and spallation typically increases with decreasing atomic number, with hydrogen being the most efficient.

The contribution to Eq.~\eqref{eq:BB} made by the density effect correction, $\delta$, is only significant for primary particles with kinetic energies that exceed their rest mass (\emph{e.g.} $\geq$ 1~Gev/n) ~\cite{Fermi1945, Sternheimer1960, Sternheimer1966,Sternheimer1982,Crispin1970}. This exceeds the energy of primary particle considered in this study and does not play a significant role in material selection. The shell correction, $C/Z_{2}$, provides a correction to the stopping power for ions $\leq$ 200 MeV/n, where their velocity is equal to or less than the orbital velocity of the lattice electrons. This correction, however, is of most consequence to ions with energies less than approximately 5~MeV/n, and thus does not play a important role in material selection.


The ionization potential, $I$, provides the largest opportunity to perturb the medium's properties in order to instigate specific changes in the emerging particle spectra that more closely model the desired field. It describes how easily a target material can absorb the kinetic energy imparted from the projectile through electronic and vibrational excitation. Unlike the density and shell corrections, whose relative contribution to stopping is strongly dependent on the projectile's energy or atomic charge, the ionization potential, $I$, is characteristic of the target material only and is independent of the properties of the projectile ion. Since the contribution of $I$ to stopping is logarithmic, small changes in its value do not produce major changes in the stopping cross section. \cite{Ziegler2010,cabrera2003} This provides an opportunity to make fine adjustments to the energies of the emerging particles by making perturbations around the measured values of the mean excitation potential for materials under consideration.

As shown in Eq.~\ref{eq:BB} and Eq.~\ref{eq:bradtandpeters}, spallation, and especially the energy loss spectrum for a heavy ion beam in a particular material, is strongly dependent on the beam species, energy, and the properties of the target material being traversed. Polymers and hydrogenated materials are favorable materials because, per unit mass, these hydrogenous materials cause higher fragmentation of high-energy heavy ions and stop more of the incident low-energy particles than other materials \cite{Zeitlin1997,Ziegler2010}. Polymers are suitable candidate materials because they have a high hydrogen content and have sufficient tensile strength for machining. For example, polyethylene (CH$_{2}$), with two hydrogen atoms and one carbon atom per monomer, is ideal for the design and construction of moderator blocks.

In order to generate a desired LET spectrum, the moderator geometry and thickness need to balance the effects of energy loss and fragmentation. A moderator geometry is chosen to correspond with the desired transmission of primary and progeny nuclei needed in the final spectrum. The desired fluence of particles required can be determined using data from, e.g., satellite measurements, IVA measurements during space missions, or from peer-reviewed models of the targeted spectrum \cite{ONeill2010}. Demonstrated in Figure \ref{fig:schematic}, each channel or cut represents a separate path the primary ions can travel through the block, where collisions between the primary ions and the moderator nuclei will result in projectile and target fragments and recoil products. Surviving primary ions continue with their initial velocity, losing energy by electromagnetic interactions. Because energy is lost and the LET depends on the inverse square of velocity, ions with sufficient range to fully traverse the moderator will emerge with higher LET \cite{zeitlin2005shielding}. The primary particles and the heavy projectile fragments represent the high-range LET components and the mid-range LET components of the GCR. The lighter fragment products will provide the contribution to the mid and low-range LET components of the GCR. The diameter, length, and material of each cut are chosen to induce specific spallation and energy loss events of the primary ion. This provides a method to selectively induce specific fragmentation and energy losses that result in the emerging field having the desired distribution of emerging ions and energies. These result in an emerging particle field mostly consisting of nuclei with a LET less than approximately 200 keV/$\mu$m.

For this work, 1 GeV/n iron ($^{56}$Fe) was chosen as the ion species and energy of the primary ion given that iron is the heaviest nuclei with significant contribution to absorbed dose in the GCR environment. The LET is approximately 150 keV/$\mu$m, which has been shown to be about the peak effectiveness for producing chromosomal damage indicative of cancer outcomes in murine models \cite{cucinotta2006cancer,durante2008heavy}. This energy and ion choice also has ranges much greater than the presumed shorter depths of the moderator block (approximately 25 cm). This ensures that the effects of fragmentation dominate while instigating a positive dose attenuation and minimal change in the LET of primaries that survive transport through the block.

The correct fluence of particles required for each layer is determined using numerical particle transport methods. Analytical prediction of the resulting particle species, their multiplicity, and corresponding energies is not possible to any high degree of accuracy. The energy loss of the primary will increases with depth and this begins to counter the expected decrease in average LET caused by fragmentation. As the primary ranges out and velocity decreases, the LET rises sharply at depths that are small compared to the mean free path for a nuclear interaction and the effects of energy loss outweigh those of fragmentation. Moderator geometry and thickness will need to balance the effects of energy loss and fragmentation. To overcome the highly-stochastic results of primary and progeny fragmentation, we  incrementally vary the geometry in each layer to quantify what material(s) and properties (e.g. length, width normal to the primary beam's path, etc.) of each layer can best produce a desired range of ions and resulting energies. The key factor in this approach is to match each layer thickness and width normal to the beam spot so that it contributes to a specific portion of the desired LET spectrum.

The final moderator block is designed so that the addition of each layer will result in a final field, $F(i,E')$, such that:

\begin{equation}\label{eq:blockfunctiontotal}
F(P,E') = \sum_{n}g_{n}(m,v)f(p,E_{i}) = G(M,V)f(p,E_{i}),
\end{equation}

\noindent where $f(p,E_{i})$ is the impinging field of the primary ion, $p$, with initial energy, $E_{i}$. The function $g(m,v)$ describes the individual layers of material $m$ and volume $v$ (e.g. length, width, height). The individual layers are summed and $G(M,V)$ describes the final moderator block material(s), $M$ and geometry, $V$. The function, $F(P,E')$, represents the resulting field of ion species, $P$, with energies, $E'$, that closely simulates the desired LET spectrum, e.g. the intravehicular LET spectrum measured on previous spaceflights. 

\begin{figure}[ht]
\centering
\includegraphics[width=1.0\columnwidth]{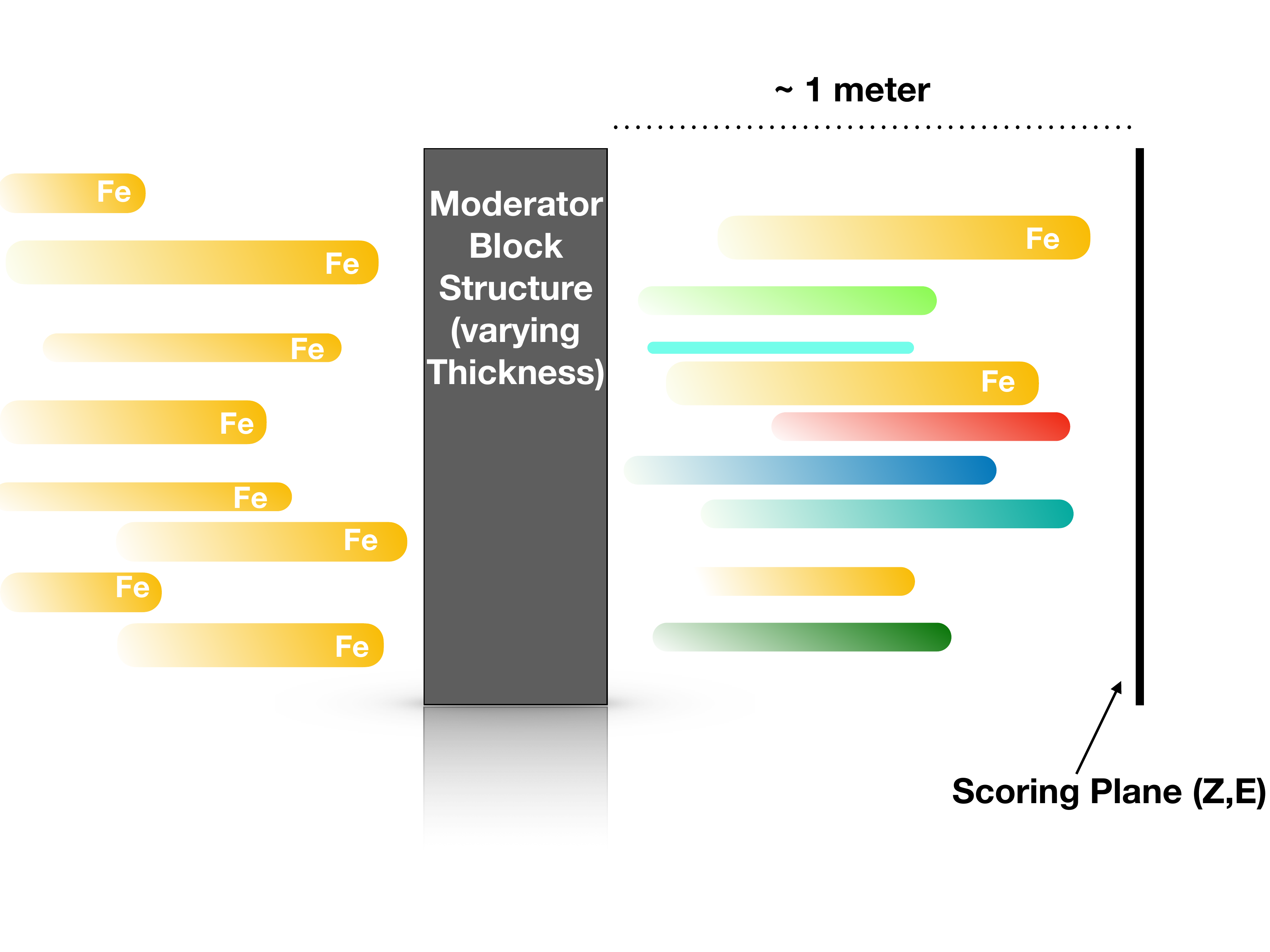} \vspace*{-0.5em}
\caption{Geometry of the Monte Carlo simulation. Iron ions enter the target from the left and surviving primaries and progeny fragments exit from the right of the block.}
\label{fig:mcmodel}
\end{figure}

A three-dimensional version of the moderator block is then recreated using combinatorial geometry for the Monte Carlo simulation. This includes accurate determinations of the width, length, and curvature of the various channels and cuts. The chemical composition and density specific to each of the moderator's layers also has to be specified for determining the material properties, such as atomic structure, ionization potential, electron shell configuration, etc. The numerical simulations are then performed using multi-core, high performance computers (HPCs) and the particle transport simulation software PHITS \cite{Niita2006}, in order to model particles traversing through thick absorbers and to approximate the desired LET spectrum.

The use of HPC (i.e. \emph{supercomputers}) allows the fast modeling of complex nuclear phenomena that would typically require significant time and computer resources. The parallelization of our numerical models allows for calculations to be distributed across multiple CPU cores and significantly increasing the number of samples sets. This greater computational power enables our models to be computed faster, since more operations can be performed per time unit, and more importantly, decrease statistical errors inherent with Monte Carlo calculations. For example, typical results were run on 5000-35,000 CPU cores over 6-12 hour time period. These produced data sets greater than approximately 2 terabytes (TB) and would equivocally take 2-3 years on a typical computer. 

PHITS features an event generator mode that produces a fully-correlated transport for all particles with energies up to $200$GeV/n \cite{Niita2006}. The software calculates the average energy loss and stopping power by using the charge density of the material and determines the momentum of the primary particle by tracking the fluctuations of energy loss and angular deviation. PHITS utilizes the SPAR code for simulating ionization processes of the charge particles and the average stopping power $dE/dx$ \cite{armstrong1973stopping,armstrong1973spar}. PHITS has been previously compared to experimental cross-section data using similar energies and materials. Zeitlin {\em et al.}~\cite{zeitlin2007fragmentation,zeitlin2008fragmentation} showed that, for large detector acceptance angles, there is good agreement between experimental beamline measurements of fragmentation cross sections and simulated outcomes that utilized PHITS to generate the expected progeny fragments and energy loss. A full description of the capabilities of PHITS and the various nuclear models utilized in the code can be found in Niita \textit{et al.} 2006.~\cite{Niita2006}.

An example two-dimensional schematic of the moderator block model used in the numerical simulation is shown in Fig.~\ref{fig:mcmodel}. The $1$GeV/n $^{56}$Fe primary beam is accelerated from the left, propagated through the moderator block, and emerges along with progeny fragments generated during spallation reactions with the block materials. The field continues to the right where a scoring plane is located $1$~meter from the moderator block face. Particle species, energy, and directional cosines are recorded for analysis and LET calculations. The LET values (in tissue) are then calculated using the stopping power formula described in Eq.~\eqref{eq:BB}. All particles are scored, including electrons, pions, neutrons, etc. However, only charged particles are considered for the final LET spectrum.

The medium traversed by the particle field emerging from the moderator is assumed to be open air. The $1$~meter distance between the back plane of the moderator and the scoring plane allows air attenuation of low-energy particles. Additionally, this space simulates experimental moderator placement with  hardware, tissue, or biological samples located down stream on the beamline.

Systematic errors are attributed to the many approximations required for a three-dimensional particle-transport Monte Carlo simulation and are, unfortunately, out of our control. The bootstrap method was utilized to determine the statistical stability of the results and minimize systematic biases in the outcomes \cite{athreya1987}.

The moderator block design was used to model IVA LET for various space exploration missions. The IVA LET spectrum measured on the U.S. Space Shuttle during the Mir Space Station Expedition 18-19 (1995) was chosen for initial validation of our approach \cite{Badhwar1998}. This particular mission was selected after identifying a rich selection of publicly accessible LET spectrums from the mission, with available measurements spanning days, weeks, and, in some cases, months. The numerical model was utilized to simulate replication of mission LET using the moderator block design. Subsequent prototype testing was performed using numerically determined model geometry. Prototype testing was performed at the NASA Space Radiation Laboratory (NSRL) at Brookhaven National Laboratory in Brookhaven, New York. A $1$GeV/n iron ($^{56}$Fe) beam was accelerated at a moderator block with beamline measurements taken using plastic scintillator detectors placed $1$~meter down the beamline to replicate model conditions. 

\begin{figure}[!ht]
\centering\includegraphics[width=1.0\columnwidth]{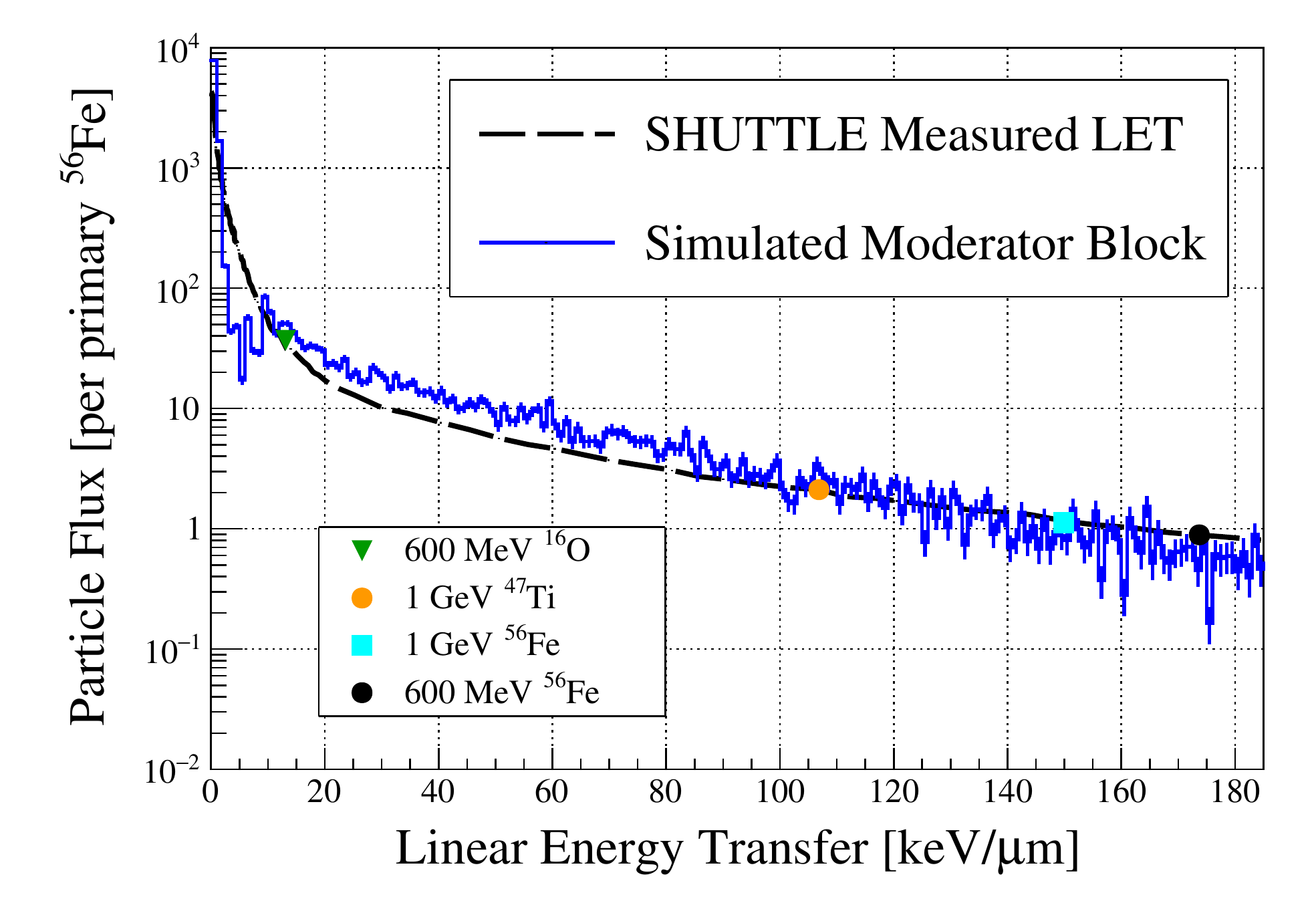}
\caption{Intravehicular particle flux versus the LET (keV/micron) field from the Shuttle-Mir 18-19 Expedition measured by Badhwar {\em et.~al.}~\cite{Badhwar1998} (dashed line), as well as the results of our moderator block model simulation (blue solid line). Note the close approximation of simulated results to the real-world curve of recorded particle flux. In addition, four single-ion exposures from current radiobiological experiments are shown (large colored symbols) to highlight the lack in breadth of energies in current radiobiological damage studies.\cite{Slaba2015,norbury2016galactic}}
\label{fig:mir}
\end{figure}

For this initial test case, every effort was made to utilize a single polymer material that is commercially available. Pragmatic decisions motivated material choice and subsequent geometry: it should be practical for a moderator block to be crafted with high precision in the machine shop of a typical accelerator laboratory, allowing for replicated use at any heavy-ion accelerator. Thus, polymers or soft materials were given priority because of sufficient tensile strength and relative ease of machining.

The results from our numerical model, along with beamline measurements from the prototype block, were compared to measurements of the U.S. Space Shuttle-Mir Expedition 18-19 IVA LET. Additional test cases of numerical models alone, without prototype correlates, were performed to include model replication of the IVA spectrum measured onboard the International Space Station (ISS) and NASA’s Orion Multi-Purpose Crew Vehicle (MPCV) during Exploration Flight Test 1 (EFT-1) in 2014.

\section*{Results}
The Mir Space Station had an orbital inclination and flight altitude of $51.6^{\circ}$ and approximately $200$ nautical miles ($370$km). Beginning in March of 1995, NASA astronauts flew several long-duration missions on the Mir Space Station, returning to earth via the U.S. Space Shuttle. Badhwar {\em et al.}~\cite{Badhwar1998} measured the integrated LET spectrum that was directly attributed to GCR ions and their spallation progeny using tissue equivalent proportional counters (TEPC) and plastic nuclear track detectors located at six different areas of the vehicle. Contributions from neutrons and non-GCR particles (e.g., Van Allen Belt ions) were not considered in model calculations in order to closely replicate real-world measured results.

\begin{figure}[ht]
\centering
\includegraphics[width=1.0\columnwidth]{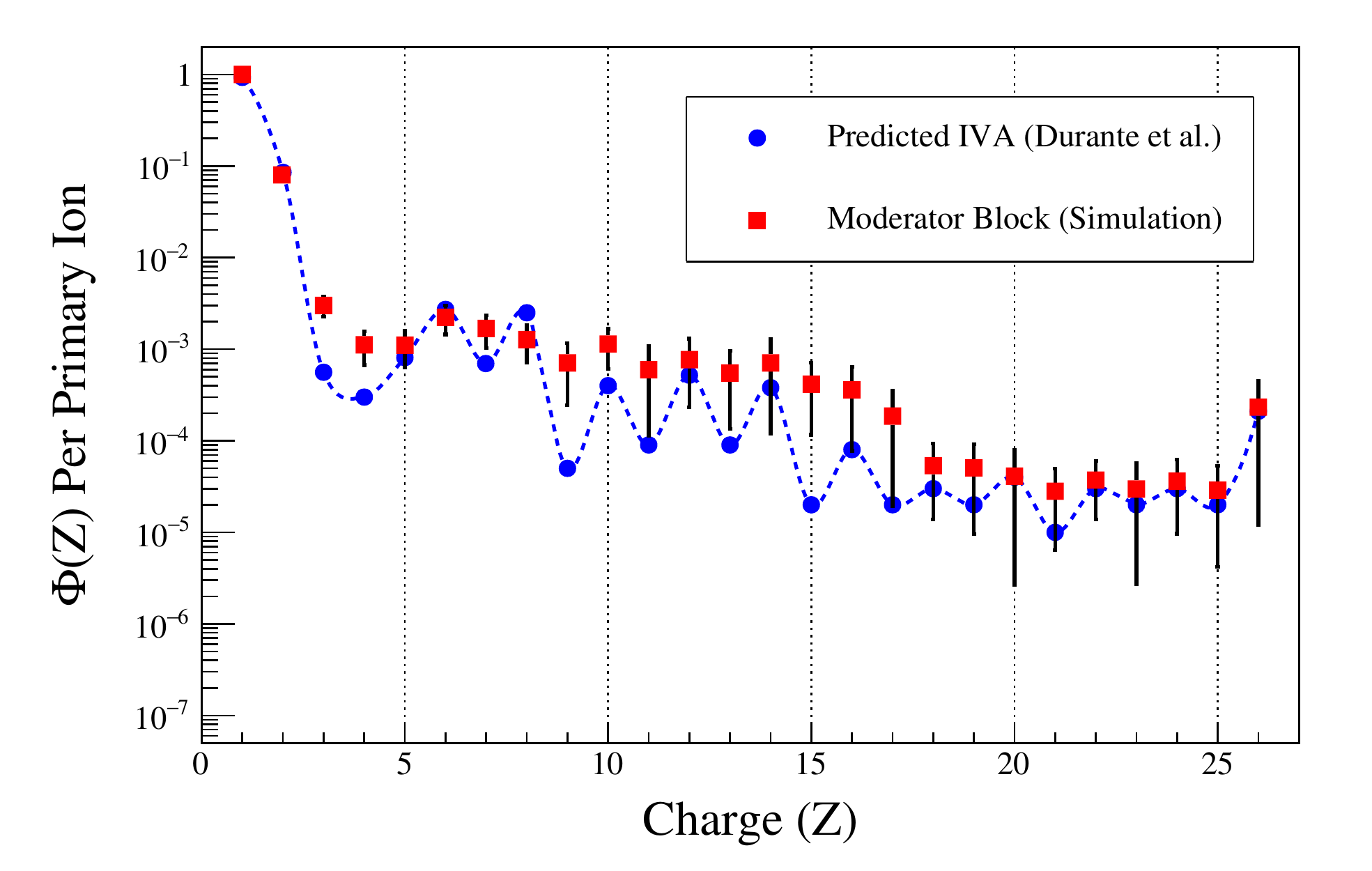}
\caption{ Comparison of predicted charge distributions. The relative abundance of intravehicular ions in the exiting field created by the moderator block (red squares) are plotted against the predicted IVA environment as published by Durante {\em et al.}~\cite{durante2008heavy} (blue circles) as a function of the atomic number $Z$. Both distributions have been normalized to the most prolific ion, hydrogen ($Z=1$).}
\label{fig:zdist}
\end{figure}

Figure \ref{fig:mir} shows the LET (per day) measured during the U.S. Space Shuttle-Mir Expedition 18-19  \cite{Badhwar1998} (black dashed line). The blue solid line represents the results of particle-transport simulations using the moderator block design developed in this work. A monoenergetic $1$GeV/n iron ($^{56}$Fe) beam passes the moderator block to output the simulated particle flux. The distribution of LET obtained from the beamline simulation demonstrates close approximation for particles having LET between $10$keV/$\mu$m and $90$keV/$\mu$m and a reasonable fit for LET up to $185$keV/$\mu$m. The output is appropriately scaled to replicate the average daily LET rate as measured during the Expedition. Note that the simulated target moderator block reproduces the spectrum over approximately five orders of magnitude. For comparison, Figure \ref{fig:mir} also identifies the individual mono-energetic ion beams (large symbols; see caption) currently used for radiobiological experiments \cite{norbury2016galactic,Slaba2015}. While the mono-energetic beams fall within the spectra of measured IVA LET, these individual ion beams do not capture the richness and diversity of the measured real-world particle flux.

We further conducted a more detailed analysis of the relative accuracy of the charge distribution resulting from the moderator block calculations compared to the predicted IVA environment as described by Durante {\em et al.}~\cite{durante2008heavy}, depicted in Figure \ref{fig:zdist}. There is a good approximation of lower-Z ions, particularly hydrogen ($Z=1$) and helium ($Z=2$), while a reasonable estimate is demonstrated for ion species $4 \leq Z \leq 26$.

\begin{figure}[ht]
\centering
\includegraphics[width=1.0\columnwidth]{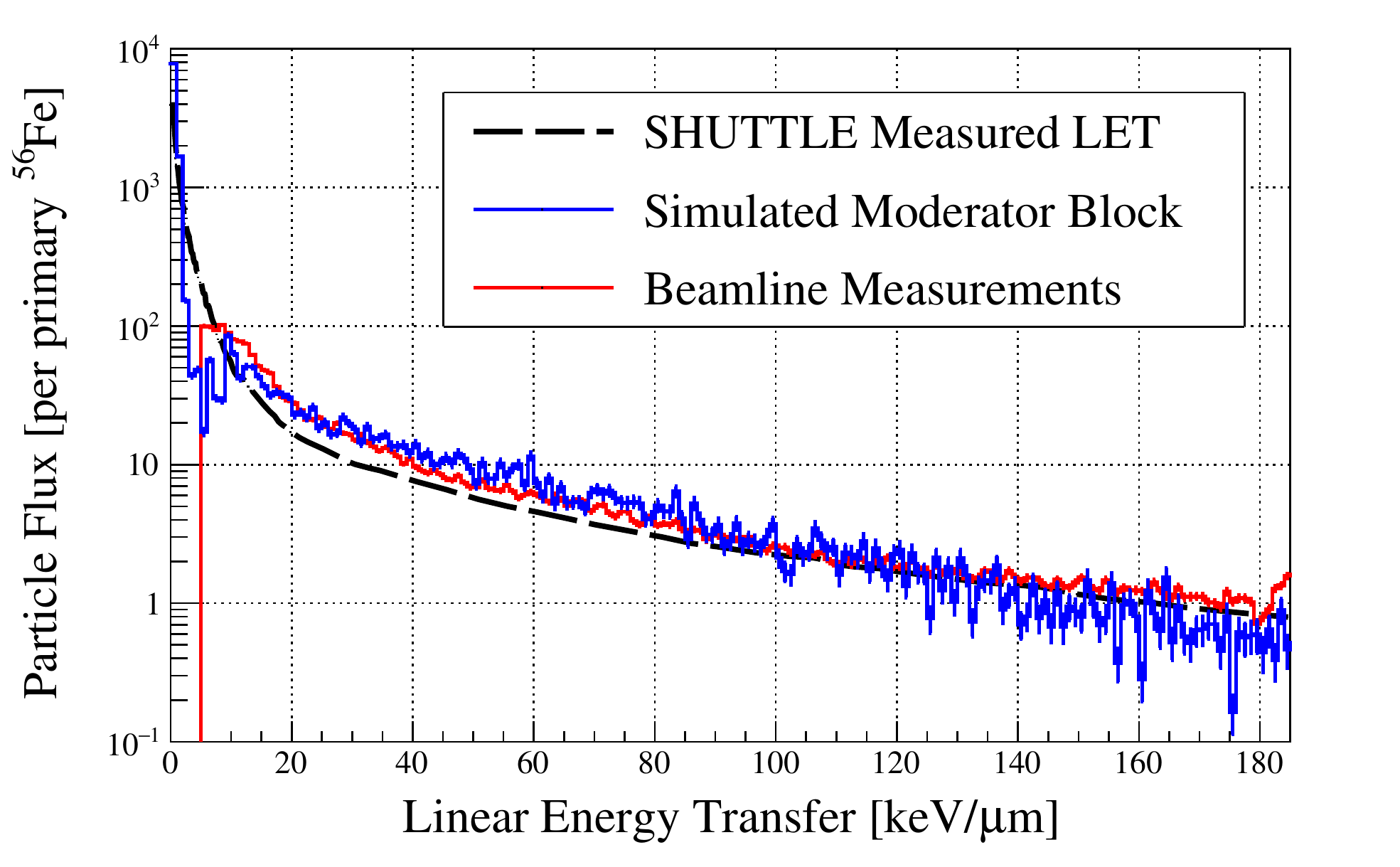}
\caption{Comparison of model with beamline measurements. The results of our moderator block model simulation (blue solid line) shown in Figure~\ref{fig:mir} are compared to beamline measurements of a prototype moderator block that replicates the numerically determined geometry (red solid line). The measured field closely matches both the measured and numerically predicted spectrum for LETs between $18$keV/$\mu$m and $185$keV/$\mu$m}
\label{fig:validation}
\end{figure}

Figure \ref{fig:validation} shows the results from beamline measurements performed at the NSRL utilizing a prototype moderator block, compared to real-world LET measurements from U.S. Space Shuttle-Mir Expedition 18-19. Spectra measured after passing through the moderator block demonstrate replication of modeled outputs as well as close approximation of real-world LET measurements from approximately $18$keV/$\mu$m to $185$keV/$\mu$m. As demonstrated in both Figure \ref{fig:mir} and Figure \ref{fig:validation}, there are some discrepancies between simulated block outputs, beamline measurements, and real-world data in the lower LET distributions. It is unclear how well the target design reproduces the LET distribution for particles with LETs approximately $5$keV/$\mu$m - $18$keV/$\mu$m. It is likely that discrepancies may be resolved by adjusting the geometry or composition of the proposed target moderators.

\begin{figure}[ht]
\centering
\includegraphics[width=1.0\columnwidth]{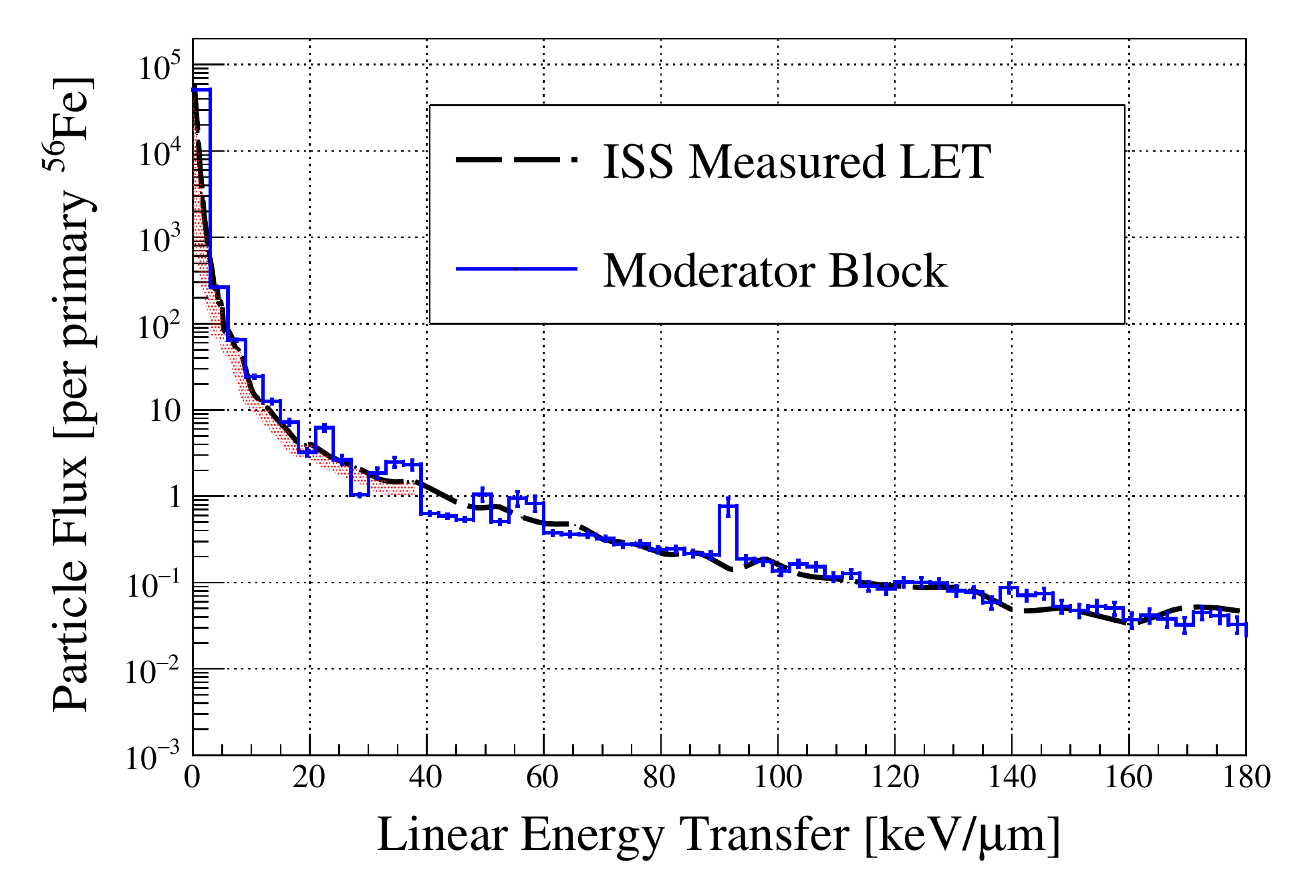}
\caption{Measured intravehicular LET field (per day) as measured onboard the ISS with the TimePix dosimeter (dashed line) compared to our moderator block model (blue solid line). The light red shading indicates uncertainty in the low-LET measurements of the TimePix dosimeter. The resulting spectrum closely replicates the real-world measured energies for continuous LET values of up to 180keV/$\mu$m over approximately seven orders of magnitude.}
\label{fig:ISS-LET}
\end{figure}

The ISS maintains an orbital inclination of $51.6^{\circ}$ and an altitude of approximately $400$km.  Figure \ref{fig:ISS-LET} shows the measured IVA LET spectrum from the ISS compared to numerical results from the moderator block design. Real-world LET measurements were taken using a TimePix hybrid pixel detector \cite{pinsky2008,stoffle2015,hoang2012let}.Note that the original LET measurements were normalized per second; here we have re-normalized to LET-per-day for consistency and for display of the estimated LET rate in units that are more relevant to radiation risk estimation for long-duration missions. The red shaded area indicates an uncertainty in the low-energy measurements (LET $\leq$ 40keV/$\mu$m) of the TimePix detector. This is most likely due to secondary electrons stopping within the instrument's silicon detector and resulting in an overestimation of approximately 10\% to  their LET values.

The measured LET spectrum in the ISS IVA environment includes all charged particles (electrons, pions, heavy charged particles, etc.). However, as with the Badhwar {\em et al.}~measurements for the U.S. Space Shuttle-Mir data presented above \cite{Badhwar1998}, onboard measurements exclude neutrons. The model spectrum approximates the real-world measured energies with reasonable accuracy for continuous LET values of up to 180keV/$\mu$m over approximately seven orders of magnitude.

The contribution of particles with low LET ($\leq$ 40 keV/$\mu$m) falls off much more slowly than what was seen for the U.S. Space Shuttle-Mir Expedition 18-19 measurements. To replicate this spectra in the moderator block design would require complex geometry, including layers with thicknesses much greater than anticipated (e.g., larger than $50$cm) that could generate the low-$Z$, high-energy particles needed to experimentally shape this portion of the LET distribution. The sharp peak in the modeled LET spectra seen at 90keV/$\mu$m result from an overabundance of ions with charges of $12 \leq Z \leq 14$ generated in the thicker portion of the moderator block. Modifications to the internal block geometry and material composition could be made to better fit dose spectra observed on the ISS without the demonstrated overabundance peaks seen in these results.
 
Measured real-world IVA spectra from NASA's EFT-1 were recently made publically available and provided an opportunity to illustrate the ability to fit the model for replication of an IVA LET spectrum from a third space vehicle \cite{kroupa2015semiconductor}. The MPCV had a flight duration of only four hours; even so, EFT-1 data are unique as the vehicle obtained a high apogee on the second orbit that included traversal through the radiation-dense Van Allen Belts and briefly into the interplanetary radiation environment. TimePix-based radiation detectors were operational shortly after liftoff and collected data for the duration of the mission \cite{Bahadori2015}.

\begin{figure}[ht]
\centering
\includegraphics[width=1.0\columnwidth]{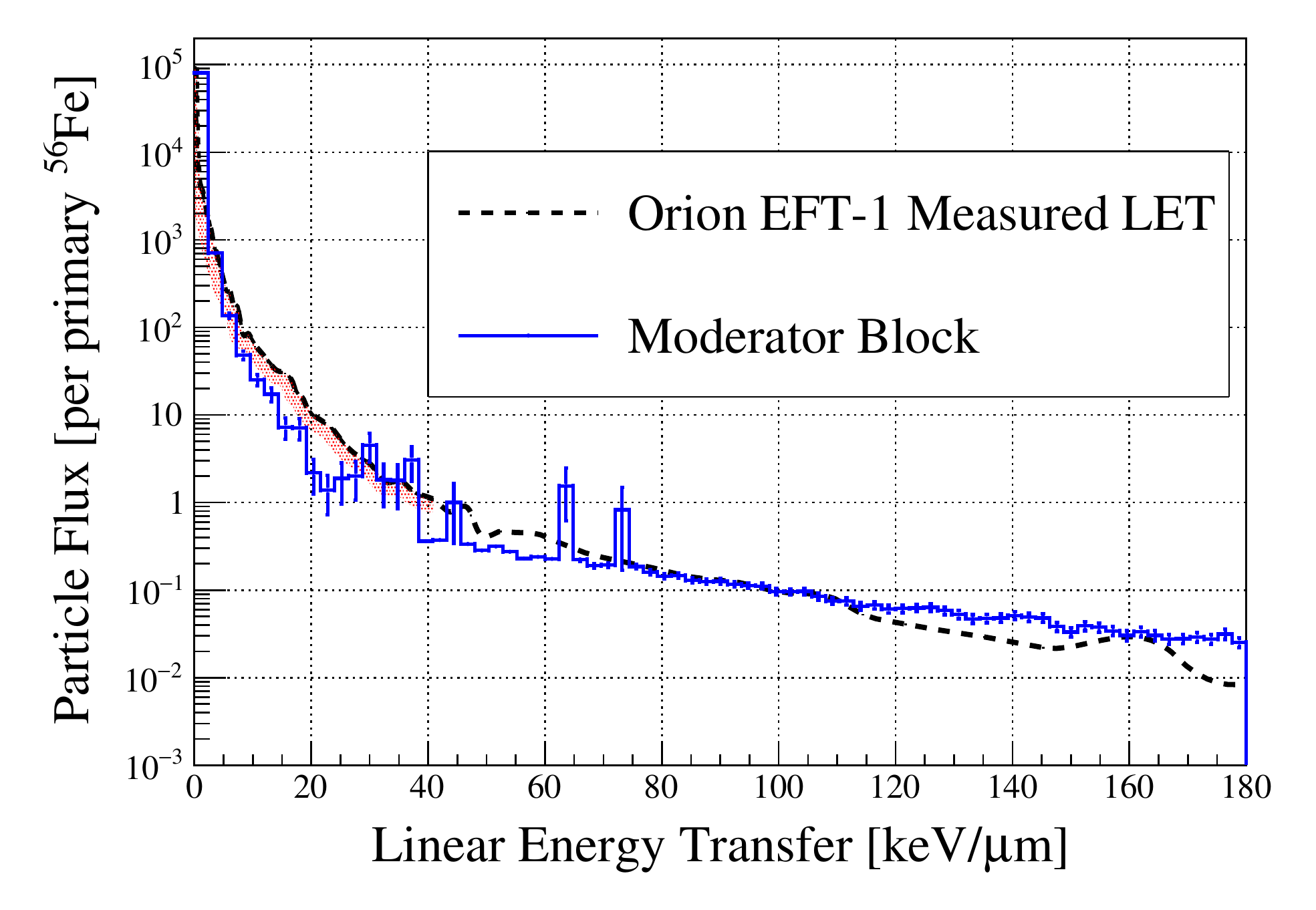}
\caption{LET field measured during the EFT-1 flight of NASA's Orion Multi-Purpose Crew Vehicle (dashed line) compared to moderator block numerical outputs. The EFT-1 mission lasted approximately four hours and included two orbits with a peak altitude of approximately $5800$km. The LET was measured for the duration of the entire flight and averaged to LET per day for this experiment. The exposure includes both interplanetary and Van Allen belt radiation fields. The light red shading indicates uncertainty in the low-LET measurements of the TimePix dosimeter. Modeled results are demonstrated by the blue solid line.}
\label{fig:eft1}
\end{figure}

EFT-1 flight data are shown in Fig.~\ref{fig:eft1} along with the results of the moderator block model modified to replicate the unique EFT-1 spectrum. While modeled results fit reasonably well compared to the flight measurements, there are visible fluctuations in the $30$-$80$keV/$\mu$m range that weakly correlates to a smaller fluctuation found from $30$-$50$keV/$\mu$m in the measured data. It is not yet clear whether these are indicators of the true nature of the measured LET spectrum, or are simple statistical fluctuations resulting from the smaller measurement period of the EFT-1 flight. Moderator layers made of polymers as thick as $100$cm are required to produce this LET spectrum. The sharp peaks at approximately $65$keV/$\mu$m and $73$keV/$\mu$m in figure \ref{fig:eft1} are due to an overabundance of ions with charge $Z \leq 6$ in block layers of $90$cm and thicker. We note that effort was made to use few hydrogen-rich materials; the presence of these peaks suggesting ion overabundance may be an indication that other low-$Z$ materials and metamaterials should be considered in future studies.

\section*{Discussion} \label{sec:discussion}
Our initial results indicate that the moderator block approach is capable of generating a complex mix of nuclei and energies that appear to more accurately simulate the space radiation environment than previous terrestrial radiation analogs. Model results show qualitative agreement with beamline measurements found in peer-reviewed literature when adapted to a geometry and environment representative of the experiment setup. This approach could enhance ground-based radiation studies by providing a more accurate recreation of the space radiation environment and allowing for a continuous generation of ionizing radiation that matches the LET spectrum and dose-rate of GCR for experimental purposes. Further, the model can be adapted to multiple scenarios, as demonstrated by simulation of varied intravehicular environments of the U.S. Space Shuttle, ISS, and MPCV vehicles. This approach could be additionally utilized to simulate the external GCR field, a planetary surface spectrum (e.g., Mars), or the local radiation environment of orbiting satellites, allowing for the characterization of multiple radiation environments that may be encountered during future space exploration. We emphasize that our model can generate both thermal and fast spallation neutron products, though these data were not included in the results presented here for more transparent comparison with measured real-world flight data. In future work, more extensive measurement of real-world IVA neutron spectrum could provide much-needed insight regarding the neutron contribution to the IVA particle spectrum and allow for higher fidelity comparisons between real-world data and model capability.

An important outcome of the results discussed here is the demonstration of validity for use of Monte Carlo numerical techniques in determining complex physical outcomes using high-performance, multi-core computers. The results presented above demonstrate computational alternatives of complex dynamics that are difficult to mimic in a laboratory setting. The recent advances in multi-core computation techniques allow for decreasing statistical errors by drastically increasing the number of samples sets. The simulation results reported for each test case required massive computation resources. Each model (U.S. Space Shuttle-Mir, ISS, EFT-1) utilized the equivalent of 135,000 cpu hours (equivalent to approximately 2.5 years of computation on a typical computer) and generated 2.5 TB of data using 5000 or more CPU cores. With the use of supercomputers, these computations were performed in roughly 10 hours. Remarkably, recent updates to the high-performance computing cluster utilized reduces the same computational time to approximately 60 minutes. The application of high-performance computational techniques allows for the adaptation of the moderator block design for high-fidelity radiation studies of materials and human health outcomes.

The ability to better simulate the space radiation environment in terrestrial research efforts has the potential to substantially enhance  understanding of the space radiation environment, allowing for rapid advances in the understanding of human radiobiological health during long-duration spaceflight. Simultaneously, such capability could drastically reduce costs and risks associated with space radiobiological research efforts by providing a high-fidelity terrestrial analog. There is a pressing need for better understanding of the true health risk imposed by the space radiation environment on future human exploration missions. The approach presented here has the potential allow rapid improvements to radiobiological studies aimed at addressing these concerns, and can additionally be generalized to other radiation spectra with wide applicability for general radiation studies of unique and extraterrestrial environments. Such a capability would fill a much-needed niche not just for radiobiological research, but also for the development of shielding, electronics, and other materials for the space environment.

\section*{Acknowledgments}
\noindent J.C.C.~would like to thank Nicholas Stoffle for the discussions about the ISS and EFT-1 LET measurement data, James
Ziegler for reviewing the concept and theory, and Serena Aunon-Chancellor for the many proof-reads
and valuable input. \newline
\newline \noindent H.G.K.~acknowledges support from the NSF (Grant No.~DMR-1151387). Part of the work of H.G.K.and J.C.C.~has been based upon work supported by the Office of the Director of National Intelligence (ODNI), Intelligence Advanced Research Projects Activity (IARPA), via Interagency Umbrella Agreement IA1-1198. The views and conclusions contained herein are those of the authors and should not be interpreted as necessarily representing the official policies or endorsements, either expressed or implied, of the ODNI, IARPA, or the U.S. Government. The U.S. Government is authorized to reproduce and distribute reprints for Governmental purposes notwithstanding any copyright annotation thereon.\newline
\newline \noindent The authors acknowledge the Texas Advanced Computing Center (TACC) at The
University of Texas at Austin for providing HPC resources that have contributed to the research
results reported within this paper.

\section*{Author contributions statement}
\noindent J.C.C.~conceived, conducted, and analyzed the experiment. J.C.C., S.B.G., and H.G.K.~provided statistical analysis of the results. R.S.B, J.R.F.~and K.A.C.~contributed to the radiobiology discussion and interpretations. All authors reviewed the manuscript.

\end{document}